\documentclass[12pt]{article}
\usepackage{amssymb}
\newcommand{\ft}[2]{{\textstyle\frac{#1}{#2}}}

\def\rmd{{\rm d}}

\newcommand{\namef}{energy}

\def\a{\alpha}
\def\b{\beta}

\def\e{\epsilon}

\def\g{\gamma}

\def\i{\iota}

\def\l{\lambda}

\def\u{\upsilon}

\def\G{\Gamma}

\chardef\tempcat=\the\catcode`\@
\catcode`\@=11
\def\cyracc{\def\u##1{\if \i##1\accent"24 i
    \else \accent"24 ##1\fi }}
\newfam\cyrfam
\font\tencyr=wncyr10
\def\cyr{\fam\cyrfam\tencyr\cyracc}

\font\cmsslll=cmss10 at 14 pt

\def\XX{{\mathsf X}}
\def\YY{{\mathsf Y}}
\def\ZZ{{\mathsf Z}}
\def\WW{{\mathsf W}}
\def\KK{{\mathsf K}}

\DeclareFontFamily{OT1}{msb}{}{}
\DeclareFontShape{OT1}{msb}{m}{n}
 {  <5> <6> <7> <8> <9> <10> gen * msbm
      <10.95><12><14.4><17.28><20.74><24.88>msbm10}{}
\DeclareMathAlphabet{\bubble}{OT1}{msb}{m}{n}

\def\bR{{\bubble R}}

\def\bC{{\bubble C}}
\def\bH{{\bubble H}}

\newfont{\goth}{eufm10 scaled \magstep1}

\def\gh{{\mbox{\goth h}}}

\def\gt{\mbox{\goth t}}
\def\gu{\mbox{\goth u}}

\def\gso{\mbox{\goth so}}
\def\gsu{\mbox{\goth su}}
\def\gspin{\mbox{\goth spin}}
\def\gsp{\mbox{\goth sp}}

\def\ghol{\mbox{\goth hol}}

\def\Sp#1{{\mathrm{Sp(#1)}}}

\def\square{\kern1pt\vbox
            {\hrule height 0.6pt\hbox{\vrule width 0.6pt\hskip 3pt
 \vbox{\vskip 6pt}\hskip 3pt\vrule width 0.6pt}\hrule height 0.6pt}\kern1pt}

\def\ra{\rightarrow}

\newtheorem{Th}{Theorem}
\newtheorem{Prop}{Proposition}
\newtheorem{Cor}{Corollary}
\newtheorem{Lem}{Lemma}
\newtheorem{Def}{Definition}

\def\bt{\begin{Th}}
\def\et{\end{Th}}
\def\bp{\begin{Prop}}
\def\ep{\end{Prop}}
\def\bc{\begin{Cor}}
\def\ec{\end{Cor}}
\def\bl{\begin{Lem}}
\def\el{\end{Lem}}
\def\bd{\begin{Def}}
\def\ed{\end{Def}}

\def\pf{\noindent{\it Proof:\ }}
\def\qed{\hfill\square}
\def\n{\nabla}
 \def\ot{\otimes}

\def\be{\begin{equation}}
\def\ee{\end{equation}}
\def\arr{\begin{array}{rlll}}
\def\ea{\end{array}}
\def\bea{\begin{eqnarray}}
\def\eea{\end{eqnarray}}
\def\bean{\begin{eqnarray*}}
\def\eean{\end{eqnarray*}}

\catcode`@=11
\@addtoreset{equation}{section}
\catcode`@=12

\def\3s{3-Sasakian manifold}

\def\qk{quaternionic-K\"{a}hler manifold}

\begin{document}
\begin{titlepage}
\begin{flushright}
KUL-TF-01/17\\
hep-th/0109094
\end{flushright}

\begin{center}
{\LARGE Flows on quaternionic-K\"{a}hler and\\[0.3cm]
very special real manifolds}
\vskip 1.0 true cm {\cmsslll Dmitri V.\ Alekseevsky$\,^{1}$, Vicente
Cort\'es$\,^{2}$,\\[0.3cm] Chandrashekar Devchand$\,^{3}$  \ and \ Antoine Van
Proeyen$\,^{4}$} \vskip 0.3 true cm
{\cyr $^{1}$ Centr ``Sofus Li'', Gen.\ Antonova 2 - 99,  117279 Moskva}\\
{\small Department of Mathematics, University of Hull,
Cottingham Road, Hull, HU6 7RX, UK \\ D.V.Alekseevsky@maths.hull.ac.uk}
\vskip 0.2 true cm
{\small $^2$ Mathematisches Institut, Universit\"at Bonn,
Beringstr. 1, D-53115 Bonn\\ vicente@math.uni-bonn.de}
\vskip 0.2 true cm
{\small $^3$ Mathematisches Institut,  Universit\"at Bonn,
Beringstr. 1, D-53115 Bonn\\ devchand@math.uni-bonn.de} \vskip 0.2 true cm
{\small $^4$ Instituut voor Theoretische Fysica, Katholieke Universiteit
Leuven\\ Celestijnenlaan 200D, B-30001 Leuven \\
Antoine.VanProeyen@fys.kuleuven.ac.be}
\vskip 0.5 true cm

{\bf Abstract}
\end{center}
{\small BPS solutions of 5-dimensional supergravity correspond to certain
gradient flows on the product $M \times N$ of a quaternionic-K\"{a}hler
manifold $M$ of negative scalar curvature and a very special real
manifold $N$ of dimension $n \ge 0$. Such gradient flows are generated by
the ``energy function'' $f = P^2$, where
$P$ is a (bundle-valued) moment map associated to $n+1$ Killing vector
fields on $M$. We calculate the Hessian of $f$ at critical points and
derive some properties of its spectrum for general quaternionic-K\"ahler
manifolds. For the homogeneous quaternionic-K\"{a}hler manifolds we prove
more specific results depending on the structure of the isotropy group.
For example, we show that there always exists a Killing vector field
vanishing at a point $p\in M$ such that the Hessian of $f$ at $p$ has
split signature. This generalizes results obtained recently for the
complex hyperbolic plane (universal hypermultiplet) in the context of
5-dimensional supergravity. For symmetric quaternionic-K\"{a}hler manifolds
we show the existence of non-degenerate local extrema of $f$, for
appropriate Killing vector fields. On the other hand, for the non-symmetric
homogeneous quaternionic-K\"{a}hler manifolds we find degenerate
local minima.
 } \vfill \hrule width 3.cm
{\small \noindent This work was supported by the priority programme
``String Theory''of the Deutsche For\-schungs\-ge\-mein\-schaft.}
\end{titlepage}
\tableofcontents{}

\section{Introduction}
Theories of 5 dimensional supergravity have recently obtained increased
attention in the
context of the AdS/CFT correspondence (for a review,
see~\cite{Aharony:1999ti}) and for a supersymmetrisation of the
Randall--Sundrum (RS) scenario~\cite{Randall:1999ee,Randall:1999vf}. In
both cases one eventually uses a 5-dimensional metric of the form
\begin{equation}
\rmd s^2 = a(x^5)^2 \rmd x^{\underline{\mu}} \rmd x^{\underline{\nu}}
\eta_{\underline{\mu\nu}}  + (\rmd x^5)^2\,, \label{5dmetric}
\end{equation}
where $\underline{\mu },\underline{\nu }=0,1,2,3$. We thus have a flat
4-dimensional space with a warp factor $a$ that depends on the fifth
coordinate $x^5$. The warp factor is interpreted as the energy scale of
the renormalization group flow in the comparison between 5-dimensional
AdS theories and 4-dimensional conformal theories. For this application
it should therefore run from a low value (infrared: IR) to a high value
(ultraviolet: UV). For the RS scenario it should have a maximum at the
value of $x^5$ which we want to associate with the position of a domain
wall, and should drop off at both infinities $x^5=\pm \infty $ towards
zero. We are then considering a scenario with one domain wall, which has
been called `smooth'. This essentially means that the configuration is a
solution of the field equations of a 5-dimensional (matter-coupled)
supergravity theory without extra sources (which would be singular
insertions of a brane in the bulk theory).

Supersymmetric theories in 5 dimensions with the minimal number of
supersymmetries (8 real supercharges) have scalars which occur in vector
multiplets and in hypermultiplets. Other multiplets, like tensor
multiplets, could be added, but would not change anything below. The
kinetic terms of the scalars define a metric on the target manifold
$M\times N$ that is a direct product of a quaternionic-K\"{a}hler manifold $M$
of dimension $4r$ of negative scalar curvature\footnote{We consider local
supersymmetry. For rigid supersymmetry, the scalar curvature of $M$ would
be zero, implying that it is (locally) a hyper-K\"{a}hler manifold.
Furthermore, we consider here always theories in a 5-dimensional space
with Minkowski signature.} parametrized by the scalars of the $r$
hypermultiplets and a very special real manifold $N$ of dimension $n$,
parametrized by the scalars of the $n$ vector
multiplets~\cite{Gunaydin:1984bi,brokensi}. The general actions have been
conveniently written down in~\cite{AnnaGianguido}. We will recall the
notion of quaternionic-K\"{a}hler manifolds in section~\ref{ss:qKmap} and of
very special real manifolds in section~\ref{ss:vspR}.

For the above-mentioned applications, one has to look for supergravity
solutions for which the only non-zero fields are the scalars and the warp
factor $a$ in (\ref{5dmetric}), and these depend only on $x^5$. The
kinematics is then determined by the kinetic terms, encoded in the
geometry of the target manifold, and by the scalar potential. In
supersymmetric theories with 8 or more real supercharges the potential is
determined by the gauging of (infinitesimal) isometries of the manifold.
An isometry can be gauged if there is a vector in the theory that can
serve as a connection. The theory contains $n+1$ vectors if $n$ is the
dimension of the very special real manifold. Indeed, pure supergravity
contains already 1 vector, the `graviphoton', while the other $n$
originate from the vector multiplets. For every isometry there is a
moment map, as we shall recall in section~\ref{ss:qKmap}. When we gauge
$n+1$ of these, the potential depends on the `dressed moment map' (see
section~\ref{ss:dressedP}), which is a linear combination of $n+1$ moment
maps on $M$ with functions on $N$ as coefficients. This is an $\gsp (1) =
\gsu (2)$ triplet $P^\alpha $. The scalars of supersymmetry-preserving
(BPS) solutions of the theory have to take values in the submanifold
determined by the condition~\cite{Ceresole:2001wi}
\begin{equation}
  \frac{\partial }{\partial \phi ^x}\left( \frac{P^\alpha }{|P|}\right)
  =0\,,
 \label{dxQ0}
\end{equation}
where $\phi ^x$, with $x=1,\ldots ,n$, are the coordinates of $N$, and
$\alpha =1,2,3$. Under this condition\footnote{This condition can be
relaxed if the 4-dimensional part of the metric is generalized from the
flat one in (\ref{5dmetric}) to a curved
one~\cite{LopesCardoso:2001rt,Chamseddine:2001ga,Chamseddine:2001hx}.},
the scalar potential depends only on the `\namef\
function'~\cite{DW,Ceresole:2001wi}
\begin{equation}
f= \ft{3}{2}W^2=P^\alpha P^\alpha\,.
 \label{deffW2}
\end{equation}
The solutions are then determined by corresponding `flow equations' which
determine the scalars as functions of $x^5$. These equations are (with
prime denoting the derivative with respect to $x^5$)
\begin{equation}
  \frac{a'}{a}=\pm  W\,,\qquad
  \phi^{x\prime}=\mp 3 g^{xy}\frac{\partial }{\partial \phi ^y} W\,, \qquad
  q^{X\prime}=\mp 3 g^{XY}\frac{\partial }{\partial q^Y} W\,,
 \label{floweqns}
\end{equation}
where $W$ is the positive root in (\ref{deffW2}), and the $\pm $ sign can
be chosen according to the sign of $a'/a$, but then has to be used
consistently in the other equations. The sign can flip when $W$ reaches a
zero. Analogous to the coordinates $\phi ^x$ for the real manifold, we
see here the coordinates of the quaternionic-K\"{a}hler manifold $q^X$
(scalars of the hypermultiplet), where $X=1,\ldots ,4r$ for $r$ the
quaternionic dimension of $M$. These equations thus determine a gradient
line on the product of the manifolds $N$ and $M$ parametrized by $x^5$,
which runs from $-\infty $ to $\infty $.

The equations imply that $(\ln a)''\leq 0$. The essential properties of
the flow can therefore be seen by analyzing fixed points of the gradient
flow and zeros of $W$. The former are the stationary values of the
scalars. We look for solutions that have `fixed points' at $x^5=\pm\infty
$. These fixed points can be found from algebraic
equations~\cite{Ceresole:2001wi}.

The behaviour of a solution near a fixed point $p$ is determined by
whether  $W$ increases or decreases when we approach $p$ in a certain
direction. This can be read off from the $(n+4r)\times (n+4r)$ matrix
\begin{equation}
   {\cal U}_\Sigma{}^ \Lambda \equiv\left.
\frac{3}{W}  g^{\Lambda \Xi}\partial_\Sigma \partial_\Xi  W
\right|_{\partial W =0}=\left. \frac{3}{2f}  g^{\Lambda \Xi}\partial
_\Sigma \partial_\Xi  f \right|_{\partial f =0} \,,
 \label{defcalU}
\end{equation}
where $\Lambda ,\Sigma $ enumerate all the scalars, and thus $\partial
_\Lambda $ contains derivatives with respect to $\phi ^x$ and $q^X$. If
the flow is along a direction corresponding to a positive part of this
matrix, the scalars flow to this fixed point with large values of the
warp factor $a$, and the point is called a UV attractor. If the flow is
along a direction where the matrix ${\cal U}$ is negative, the scalars
are attracted to this point for small values of $a$, and this point is
called an IR attractor or IR fixed point (see
e.g.~\cite{Behrndt:2001qa,Ceresole:2001wi} for more details). The
eigenvalues of ${\cal U}$ are also the conformal weights of the
corresponding operators in the conformal dual to the supergravity theory.

The purpose of this paper is to derive general properties of such flows.
These are mostly determined from knowledge of the matrix ${\cal U}$. Our
main result is a suitable formula for this matrix. Furthermore we can
derive general results on the possibility of UV and IR critical points in
symmetric or homogeneous quaternionic-K\"{a}hler manifolds.

The paper is organized as follows. In section~\ref{ss:qKmap} we recall the
basis properties of qua\-ter\-ni\-onic\--K\"{a}hler manifolds and the moment
map. We then analyse the part of the matrix ${\cal U}$ (Hessian of the
\namef) for a pure quaternionic-K\"{a}hler manifold in
section~\ref{ss:Hessianq}, and derive properties of attractor points. The
very special real manifolds are introduced in section~\ref{ss:vspR} and
the adapted moment map in section~\ref{ss:dressedP}. This allows us to
find the properties of the full Hessian in section~\ref{ss:Hessfull}.
Finally we give conclusions in section~\ref{ss:conclusions}. An effort is
made to translate mathematical formulae in notation readable to
physicists and vice versa. In particular, we have given a presentation of
very special real geometry which is accessible to mathematicians. Some
remarks on our notation are gathered in the appendix.

\section{Quaternionic-K\"{a}hler moment map} \label{ss:qKmap}
We start by recalling the notion of quaternionic-K\"{a}hler manifold. Let
$(M,g)$ be a Riemannian manifold. A {\bf quaternionic-K\"{a}hler structure}
$Q$ on $M$ is a rank 3 subbundle $Q \subset {\rm End}(TM)$ invariant
under parallel transport such that locally $Q = {\rm span}\{ J_1,J_2,J_3
= J_1J_2\}$, where the $J_{\alpha}$
are locally defined skew-symmetric almost complex structures on $M$.

With respect to local coordinates $q^X$, with $X = 1,\ldots 4r = \dim M$,
the almost complex structures $J_{\alpha}$ have components $J_{\alpha
X}{}^Y$ satisfying\footnote{Here and below, a sum over repeated indices
is understood.}
\begin{equation}
  J_{\alpha X}{}^YJ_{\beta Y}{}^Z=-\delta _{\alpha \beta }\delta _X{}^Z
  +\varepsilon _{\alpha \beta \gamma }J_{\gamma X}{}^Z\, ,
 \label{JJcomp}
\end{equation}
where $\varepsilon _{\a\b\g}$ is completely antisymmetric with
$\varepsilon _{123}{=}1$. The invariance of $Q$ under the Levi-Civita
connection $\n$ is tantamount to the existence of a triplet of one-forms
$\omega _\alpha$ such that
 \be
    \nabla J_\alpha =-2\varepsilon _{\alpha \beta \gamma }\omega _\beta J_\gamma
    \, .
 \label{DJ0}\ee
The $\omega _\alpha$ may be determined from this equation, which can be
rewritten as the full covariant constancy of the complex structures; in
components:
\begin{equation}
   {\cal D}_Z J_{\alpha X}{}^Y := \partial_ZJ_{\alpha X}{}^Y-\Gamma_{ZX}^W
  J_{\alpha W}{}^Y+ \Gamma _{ZW}^YJ_{\alpha X}{}^W
   +2\varepsilon _{\alpha \beta \gamma } \omega _{\beta Z}J_{\gamma
   X}{}^Y=0\, .
 \label{calD}
\end{equation}
Here $\Gamma$ are the Christoffel symbols of $\n$ and
$2\varepsilon_{\alpha\beta \gamma}\omega_\b$ is the connection matrix of
the connection induced by $\n$ in the bundle $Q$. We note that $\cal D$
is here a connection in $Q \otimes \bR^3$ induced by the Levi-Civita
connection on $Q$ and the connection on the trivial rank 3 bundle over $M$
defined by ${\cal D} e_\a = 2\varepsilon_{\alpha\beta \gamma}\omega_\b
e_\g$, where $e_\a$ is the standard basis of $\bR^3$. Note that equation
(\ref{calD}) means that the section $J := J_\a \otimes e_\a$ is parallel
with respect to $\cal D$.

In general, ${\cal D}$ contains the gauge field
of all the transformations of the object on which it acts (see appendix).

A Riemannian manifold admits a quaternionic-K\"{a}hler structure if and only
if its holonomy group is a subgroup of $\Sp{r}\Sp{1}$, where $4r = \dim
M$. The group $\Sp{r}\Sp{1}$ is the linear group normalizing a
quaternionic structure on $\bR^{4r}$ and preserving a compatible
Euclidean scalar product.

A {\bf quaternionic-K\"{a}hler manifold} of $\dim M = 4r > 4$ is a Riemannian
manifold endowed with a quaternionic-K\"{a}hler structure. In dimension 4
(the case $r = 1$) this definition would correspond simply to the notion
of oriented Riemannian 4-fold. Instead we will assume in addition that
$Q$ annihilates the curvature tensor of the manifold $(M,g)$, i.e.
\begin{equation}
  J_{\alpha X}{}^VR_{VYZW} + J_{\alpha Y}{}^VR_{XVZW}
  +J_{\alpha Z}{}^VR_{XYVW}
  +J_{\alpha W}{}^VR_{XYZV}=0\,.
 \label{QanniR}
\end{equation}
This condition is automatically satisfied if $r>1$. Then the following
result holds in all dimensions \cite{Alekseevsky1968}.

\bt \label{structureThm} The curvature tensor $R$ of a \qk\, of dimension
$4r$ is of the form
\[ R = \nu R_0 + {\cal W} \, ,\]
where $R_0$ is the curvature tensor of the quaternionic projective space,
$\nu = \frac{scal}{4r(r+2)}$ is the reduced scalar curvature and ${\cal
W}$ is an algebraic curvature tensor of type $\gsp (r)$ (the ``Weyl
curvature'').
 \et
This means that the components can be written as
\begin{eqnarray}
R_{XYZW}&=& \nu \left[ \ft12 g_{Z [X }g_{Y ]W }
              +\ft12 J^\alpha_{X Y }J^\alpha _{Z W }
              -\ft12 J^\alpha _{Z [X }J^\alpha _{Y ]W }\right]
\nonumber\\[5pt]
&& +f_X ^{iA}f_Y ^{jB}\varepsilon _{ij}f_Z ^{kC}f_W ^{\ell D}
\varepsilon_{k\ell }\Sigma_{ABCD}\,, \label{Rdecomp}
\end{eqnarray}
where the antisymmetrization of a tensor $T_{XY}$ is defined as
$T_{[XY]}:= \frac12(T_{XY}-T_{YX})$, the $f_X^{iA}$ are the vielbeins of
the manifold ($A=1,\ldots ,2r$ and $i,j=1,2$) and $\Sigma _{ABCD}$ is
completely symmetric. We do not use vielbeins in this paper, but the
interested reader may find a discussion of \qk\ in terms of vielbeins in
the article on quaternionic-K\"ahler manifolds in~\cite{ENCYCL} and a
definition of quaternionic-K\"{a}hler manifolds that starts from vielbeins
has been given in~\cite{VanProeyen:2001ng}, see
also~\cite{Fre:2001jd,D'Auria:2001kv,Ceresole:2001wi}.

To avoid confusion we emphasize that here and in all coordinate
expressions we use $X, Y, \ldots$ to denote indices from $1,\ldots, 4r$,
following the convention in the 5d supergravity literature. In coordinate
free formulas the same letters, in sans-serif font, $\XX, \YY, \ldots$,
will denote vector fields.

It follows from Theorem \ref{structureThm} that quaternionic-K\"{a}hler
manifolds are Einstein, in fact, $Ric = \nu (r+2) g$. Only in the case
$\nu = 0$, can we choose the three local almost complex structures
spanning the quaternionic structure $Q$ to be parallel and, in particular,
integrable. This case corresponds to locally hyper-K\"{a}hlerian manifolds
and is excluded in the following discussion. So from now on $\nu \neq 0$.
Supergravity fixes $\nu =-1$, but we will keep $\nu $ general below.

To any \qk\, we can associate the parallel 4-form
\[ \Omega :=
\rho_{\alpha}\wedge \rho_{\alpha}\, ,\] where the $\rho_{\a} = g(\cdot ,
J_\a \cdot )$ are the $2$-forms (``K\"{a}hler forms'') associated to any
choice of three local almost complex structures $(J_1,J_2,J_3 = J_1J_2)$
spanning $Q$ (its components are just the components of the almost
complex structures with indices lowered by the metric). Let $\KK$ be a
Killing vector field on a \qk\, $(M,g,Q)$. Then $\KK$ normalizes $Q$.
Indeed, if $(M,g)$ is locally symmetric, then the Lie algebra of all
Killing vector fields is well known. If $(M,g)$ is not locally symmetric
and $\dim M = 4r > 4$ then the holonomy Lie algebra is $\ghol = \gsp (r)
\oplus \gsp (1)$ and any Killing vector field normalizes the holonomy Lie
algebra and in particular its $\gsp (1)$-factor, which defines the
quaternionic-K\"{a}hler structure $Q$. This proves that $\KK$ normalizes $Q$,
i.e.\footnote{Here ${\cal L}_{\KK}$ is the Lie derivative, i.e.\ ${\cal
L}_{\KK} \XX = [\KK,\XX ]$ for all vector fields $\XX$.} ${\cal L}_{\KK}
J^\alpha= b^{\alpha\beta}J^\beta$ for some $b^{\alpha \beta }(q^X)$. This
amounts to
\begin{equation}
  ({\cal D}_{[X}\KK^Z)J_{Y]Z}^\alpha
  = -\nu \varepsilon^{\alpha \beta \gamma} J^\beta_{XY}P^\gamma \,,
 \label{KnormQ2}
\end{equation}
for some $P^\gamma(q^X)$ with normalization chosen for later convenience.

The 3-form obtained from the 4-form $\Omega$ by contraction,
\[ \iota_{\KK} \Omega =
 \rho_\a (\KK, \cdot) \wedge \rho_\a \]
is closed:
\[
 \rmd\iota_{\KK} \Omega = {\cal L}_{\KK}\Omega - \iota_{\KK} \rmd\Omega = 0\, .
\]

\bp The three-form  $\iota_{\KK} \Omega$ is exact
\[
 \nu \i_{\KK}\Omega = \rmd\rho \, ,\quad  \rho :=
 \langle \n \KK, J_\a\rangle \rho_\a  \, ,
\]
where $\langle \cdot , \cdot  \rangle$ is the canonical
scalar product on ${\rm End} \, TM$ normalized such that
$\langle J_\a ,J_\b \rangle = \delta_{\a \b}$.
\ep

\pf Let us compute $\rmd\rho = {\rm alt} \n \rho$ :
 \be \label{nablarhoEqu}\n \rho =
 \langle \n^2 \KK, J_\a \rangle \rho_\a
 -2 \varepsilon _{\alpha \beta \gamma }\left(
 \langle \n \KK, \omega _\b J_\gamma \rangle \rho_\a
 + \langle \n \KK, J_\a \rangle \omega_\b \otimes \rho_\gamma
 \right)  \, .
 \ee
Here, the last two terms on the right hand side cancel each other. The
first term is computed using the following lemma.

\bl Let $\KK$ be a Killing vector field on a Riemannian manifold with
curvature tensor $R$. Then the second covariant derivative of $\KK$ is
given by\footnote{In equations as the one below, where vectors are not
explicitly written, they should be understood as appearing consistently
from left to right, e.g. below: $\nabla^2_{\XX,\YY} \KK := \nabla_{\XX}
\nabla_{\YY} \KK -\nabla_{\nabla_{\XX} \YY} \KK = R(\XX , \KK)\YY$. Using
the covariant derivative $\cal D$ instead, this equation may be written
in a coordinate basis in the form $ {\cal D}_X{\cal D}_Y \KK^Z =
R{}^Z{}_{YXW}\KK^W\,$.}
 \be
\nabla^2 \KK = R(\cdot , \KK)\, .
 \ee \el

\pf We prove first that the tensor $\nabla^2 \ZZ - R(\cdot , \ZZ)$ is
symmetric for any vector field $\ZZ$. This follows from the Bianchi
identity:
\[
 \nabla^2_{\XX,\YY} \ZZ - R(\XX, \ZZ)\YY
 - (\nabla^2_{\YY,\XX} \ZZ - R(\YY, \ZZ)\XX) =
R(\XX,\YY)\ZZ - R(\XX, \ZZ)\YY + R(\YY, \ZZ)\XX  = 0\, .
\] So it is
sufficient to check that $\langle \nabla^2_{\XX,\XX}\KK, \YY\rangle =
\langle R(\XX,\KK)\XX, \YY\rangle$ for all vector fields $\XX$ and $\YY$.
We can assume that $[\XX,\YY] = 0$, since we are checking an identity
between tensors; and we use the Killing equation $\langle \n_{\XX}
\KK,\YY\rangle = - \langle \n_{\YY}\KK, \XX\rangle$.
\begin{eqnarray*}
\langle \nabla^2_{\XX,\XX}\KK, \YY\rangle &=&
 \langle \nabla_{\XX}\n_{\XX} \KK - \n_{\n_{\XX} \XX}\KK ,\YY\rangle
 \\
&=& \XX  \langle \n_{\XX}  \KK,\YY\rangle -  \langle \n_{\XX}
\KK,\n_{\XX}\YY\rangle
+ \langle \n_{\YY}\KK,\n_{\XX}\XX\rangle
\\
&=& -\XX\langle \n_{\YY} \KK,\XX\rangle
 -  \langle \n_{\XX}  \KK,\n_{\XX}\YY\rangle +\langle
  \n_{\YY}\KK,\n_{\XX}\XX\rangle
  \\
&=& -\langle \n_{\XX}\n_{\YY} \KK,\XX\rangle
 -  \langle \n_{\XX} \KK,\n_{\XX}\YY\rangle\\
&=& -\langle \n_{\XX}\n_{\YY} \KK,\XX\rangle
     - \langle \n_{\XX} \KK,\n_{\YY}\XX\rangle\\
&=& -\langle \n_{\XX}\n_{\YY} \KK,\XX\rangle
     + \langle \n_{\YY}\n_{\XX} \KK,\XX\rangle\\
&=& -\langle R(\XX,\YY)\KK,\XX\rangle = \langle R(\XX,\KK)\XX,\YY\rangle
\end{eqnarray*}
\qed

\noindent

Now, using this lemma  we obtain
\begin{eqnarray} \label{nablarhoEqu2}\n \rho &=&
 \langle \n^2 \KK, J_\a \rangle \rho_\a =
  \langle R(\cdot ,\KK), J_\a\rangle \rho_\a \nonumber\\
&=& \ft{\nu}{2}\rho_\a(\cdot , \KK) \otimes \rho_\a \, .
\end{eqnarray}
Here, we use  the fact that the $\gsp (1)$-part of $R = \nu R_0 + {\cal
W}$ is given by the middle term of the first line of (\ref{Rdecomp}) (the
other terms annihilate under multiplication with $J^{\beta XY}$):
 \be
R^{\gsp (1)} = \nu R^{\gsp (1)}_0 = \ft{\nu}{2} \rho_\a J_\a \, ,
\label{sp1curv}
 \ee
see Theorem~\ref{structureThm}. For the exterior derivative we thus
obtain,
 \be
  \rmd\rho = \frac{\nu}{2}\rho_\a(\cdot , \KK)
\wedge \rho_\a = \nu \i_{\KK}\Omega\,,
 \ee
proving the proposition. \qed

\noindent
The {\bf moment map} associated to the Killing vector $\KK$ is
the section $P =  P^\a J_\a \in \G (Q)$  related to the two form $\rho$ by
$\rho = \nu g(\cdot , P)$, cf.\ \cite{Galicki:1986vh}. It follows that
 \be
\label{PEqu} P^\a = \epsilon \langle \n \KK, J_\a\rangle  \, , \quad \e :=
\ft{1}{\nu}\, .
 \ee
This is consistent with the use of $P^\alpha $ in (\ref{KnormQ2}). We are
interested in the gradient flow generated by the function
 \be
 f := P^2 := \langle P, P\rangle = \e^2\sum \langle \n \KK ,
J_\a \rangle^2\,,
 \ee
which we call the {\bf \namef}.

\bp \label{firstProp}
 The covariant derivative of the moment map $P$ is
given by
\be \label{nablaP}
 \n P = \ft{1}{2} \rho_\a (\cdot , \KK)
\otimes J_\a \, .
\ee
 The gradient of the \namef\ is
  \be {\rm
grad}\, f = P\KK = P^\a J_\a \KK = \epsilon \langle \n \KK, J_\a\rangle
J_\a \KK\, . \ee \ep

\pf The formula (\ref{nablaP}) is an immediate consequence of
(\ref{nablarhoEqu2}). Using (\ref{PEqu}) and (\ref{nablaP}) we compute
the differential $\rmd f$: \be \label{dfEqu}
 \rmd f = 2 \langle \n P, P\rangle =
P^\a \rho_\a(\cdot , \KK) = \epsilon \langle \n \KK, J_\a\rangle
\rho_\a(\cdot , \KK)\, .
 \ee
This implies the formula for the gradient. \qed

\bc The set of critical points of $P$ is
 \be
  {\rm Crit} (P) = \{ \KK = 0\} \, .
 \ee
The set of critical points of the \namef\ $f$ is the union
 \be
 {\rm Crit}(f) = \{ \KK = 0\} \cup \{ f = 0\} \, .
 \label{Critf}
 \ee \ec

The formula (\ref{nablaP}) appears in supergravity as the definition of
the moment map or `prepotential'\footnote{Note that the form
   $\rho _\alpha (\cdot ,\KK)= g(\cdot , J_\a\KK)$ has components
    $-J^\alpha _{XY}\KK^Y$.} in the component form:
\begin{equation}
 -\nu  {\cal D}_XP^\alpha ={\cal R}^\alpha_{XY}\KK^Y
 =\frac{\nu}{2}J^\alpha_{XY}\KK^Y\,.
 \label{defPSG}
\end{equation}
In supergravity $\nu =-1$. Here,
\begin{equation}
{\cal R}^\alpha_{XY}  = 2\partial_{[X}\omega^\alpha  _{Y]}+2\omega
_X^\beta \omega _Y^\gamma \varepsilon _{\alpha \beta \gamma }\,,
 \label{defcalR}
\end{equation}
are the components of the $\gsp (1)$ curvature (\ref{sp1curv}). They are
clearly proportional to the K\"{a}hler forms $\rho_\a$, yielding the formula
for the gradient
\begin{equation}
  \partial_Xf=-P^\alpha J^\alpha _{XY}\KK^Y\,.
 \label{gradfcomp}
\end{equation}

{\bf Remark:} The set $\{ \KK = 0\}$ is a union of {\it totally geodesic}
submanifolds. This follows from the fact that a connected component of the
fixed point set of a group of isometries is totally geodesic since
isometries transform geodesics to geodesics and there exists a unique
geodesic through two sufficiently close points.  If $(M,g)$ is complete
and has non-positive sectional curvature (e.g.\ if $(M,g)$ is a symmetric
space of non-compact type or, more generally, a Riemannian manifold
covered by such a space) then $\{ \KK = 0\}$ is {\it connected} since in
the universal covering of $M$ any two points are joint by a unique
geodesic.

This generalizes to any symmetric space allowed in supergravity (which
has to be non-compact due to the $\nu =-1$ condition) the result found in
the toy model (universal hypermultiplet) in~\cite{Ceresole:2001wi}.
Namely, if there is an isolated critical point, then there are no other
critical points, or, if there are two critical points then, as explained
above, they are connected by a geodesic which consists of critical points.

\section{The Hessian of the \namef\ at a critical point} \label{ss:Hessianq}
In this section we compute the Hessian of the \namef\  $f$ at critical
points and study its spectrum. For this we need the following lemma.
 \bl \label{nabla2PLemma} The second covariant derivative of the moment
map is given by:
 \be
  \n^2_{\XX,\YY}P = \ft{1}{2} \rho_\a (\YY , \n_{\XX}\KK)J_\a\, . \ee
 The Hessian of the \namef\  is:
  \be \label{HessEqu}
{\rm Hess}_f (\XX,\YY) := \n^2_{\XX,\YY} f =  P^\a \rho_\a (\YY, \n_{\XX}
\KK) + \ft{1}{2}  \rho_\a (\XX, \KK)\rho_\a (\YY, \KK)\,.
 \ee \el

\pf Using (\ref{nablaP}) we compute
\begin{eqnarray*} \n^2_{\XX,\YY}P  &=&
\ft{1}{2}g(\YY, \n_{\XX}(J_\a)\KK) J_\a + \ft{1}{2}g(\YY, J_\a \n_{\XX}
\KK) J_\a +
\ft{1}{2}g(\YY, J_\a \KK) \n_{\XX} J_\a
\\
&=& \ft{1}{2} \rho_\a (\YY , \n_{\XX} \KK)J_\a -\varepsilon _{\alpha
\beta \gamma } \left[\omega _\b(\XX) g(\YY ,J_\g \KK)  J_\a
+ g(\YY , J_\a \KK) \omega_\b(\XX) J_\g \right]  \\
&=& \ft{1}{2} \rho_\a (\YY , \n_{\XX} \KK)J_\a\,.
\end{eqnarray*}
For the Hessian of $f = P^2$ we get:
\begin{eqnarray*}
\n^2_{\XX,\YY} f &=&
2\langle \n^2_{\XX,\YY} P, P\rangle +
2 \langle \n_{\XX} P, \n_{\YY} P\rangle\\
&=& \epsilon \rho_\a (\YY, \n_{\XX} \KK)\langle \n \KK, J_\a\rangle +
\ft{1}{2} \rho_\a (\XX, \KK)\rho_\a (\YY, \KK)\\
&=&   P^\a \rho_\a (\YY, \n_{\XX} \KK) + \ft{1}{2}  \rho_\a (\XX,
\KK)\rho_\a (\YY, \KK)
\end{eqnarray*}
\qed

\noindent
Let us decompose the operator
 \be
  L_{\KK} := \n \KK =  \langle \n \KK , J_\a
\rangle J_\a + \bar{L}_{\KK} = \nu  P^\a J_\a  + \bar{L}_{\KK} \, .
\label{decompNK}
 \ee
Then $\bar{L}_{\KK}$ is a skew symmetric operator commuting with $Q$. This
follows from the fact that $\n$ and $\KK$ preserve $Q$ using the formula
 \be
  \n \KK = \n_{\KK} - {\cal L}_{\KK} \, .
  \ee
The operators $\langle \n \KK , J_\a \rangle J_\a$ and $\bar{L}_{\KK}$ are
called the $\gsp (1)$-part and the $\gsp (r)$-part of $L_{\KK}$,
respectively. The important properties of $\bar{L}_{\KK}$ are
\begin{equation}
\left(S_{\KK} \right)_{\alpha\,XY } :=J_{\alpha\, X}{}^Z
\left(\bar{L}_{\KK} \right)_{ZY} =\left(S_{\KK} \right)_{\alpha\,YX }
\,,\qquad \left(S_{\KK} \right)_{\alpha\,X }{}^X=0\,.
 \label{propcalJL}
\end{equation}

\bt The set of critical points of the \namef\  $f$ is given in
(\ref{Critf}). At a point $p\in M$ where $f=0$ the Hessian of $f$ is
given by
 \be
  {\rm Hess}_f (\XX,\XX) = \ft{1}{2} \sum \rho_\a (\XX, \KK)^2
 \ee
and hence it is positive semi-definite. Its kernel is
 \be
 {\rm ker}\,
{\rm Hess}_f = {\rm span} \{ J_1\KK,J_2\KK,J_3\KK\}^\perp \subset T_pM \,.
 \ee
At a point where $\KK=0$ the Hessian is given by
 \bea
{\rm Hess}_f (\XX,\XX) &=&   P^\a \rho_\a (\XX, \n_{\XX} \KK)
\nonumber\\
 &=& - \nu f
g(\XX,\XX) + g(\XX, S\XX)\, ,
\eea
where $S$ is the symmetric operator
 $S := P\bar{L}_{\KK} = P^\a J_\a \bar{L}_{\KK}
 =P^\alpha \left(S_{\KK} \right)_{\alpha} $.
 \et

\pf This follows immediately from (\ref{dfEqu}), (\ref{HessEqu}) and the
decomposition of $L_{\KK}$ into its $\gsp (1)$ and $\gsp (r)$-parts. \qed

\noindent
As this is a main result of this section, we give also its
component form:
\begin{equation}
\left.  \partial_X\partial_Y f\right|_{\KK=0}
  =-\nu\, f\, g_{XY}+ P^\alpha
\left(S_{\KK} \right)_{\alpha XY}\,.
 \label{HessK0}
\end{equation}

Recall that the Hesse operator ${\rm H}_f$ is defined by $g({\rm
H}_f\XX,\YY) = {\rm Hess}_f (\XX,\YY)$. We are now in a position to prove:
\bt At a point where $\KK=0$ there exists an eigenbasis for the Hesse
operator ${\rm H}_f$ of the form
\[ e_1\,,\, J_1e_1\,,\, e_2\,,\,J_1 e_2\,,\ldots ,\, e_r\,,\, J_1e_r\,;\,
 J_2e_1\,,\, J_3e_1\,,\,
J_2e_2\,,\, J_3e_2\,,\ldots ,\, J_2e_r\,,\, J_3e_r\, . \]
 The corresponding eigenvalues are
\[ \l_1 - \nu f\,,\, \l_1 - \nu f\,,\,  \l_2 - \nu f\,,\, \l_2 - \nu f\,,
 \ldots ,\,
\l_{r} - \nu f \,,\, \l_{r} - \nu f\, ; \]
\[ -\l_1 - \nu f\,,\, -\l_1 - \nu f\,,\, -\l_2 - \nu f\,,\,
-\l_2 - \nu f\,,\ldots ,\, -\l_{r} - \nu f\,,\, -\l_{r} - \nu f\, .
\] \et

\pf The eigenvectors of ${\rm H}_f = -\nu f \, {\rm id} + S$ coincide
with the eigenvectors of the operator $S = P^\a J_\a \bar{L}_{\KK}$.
Without loss of generality we can assume that $P^\a J_\a$ is proportional
to $J_1$: $P^\a J_\a = c J_1$, $c\in \bR$. Then the operator $S = P^\a
J_\a \bar{L}_{\KK} = c J_1\bar{L}_{\KK}$ commutes with $J_1$ and
anticommutes with $J_2$ and $J_3$. Let $v$ be an eigenvector of $S$.  Then
\[ Sv = \l v\, \, \quad S J_1v = J_1 Sv = \l J_1 v \, , \quad
SJ_2 v = - J_2 Sv = -\l J_2 v\, ,\quad \mbox{and} \quad
SJ_3 v = -\l J_3 v\, .
\]
Now one can easily prove by induction that there is an
eigenbasis of $S$ of the form
\[ e_1, J_1e_1, e_2,J_1 e_2, \ldots , e_r, J_1e_r, J_2e_1, J_3e_1,
J_2e_2, J_3e_2, \ldots , J_2e_r, J_3e_r \] with eigenvalues
\[ \l_1, \l_1, \l_2, \l_2,\ldots , \l_r, \l_r, -\l_1, -\l_1,
\ldots , -\l_r, -\l_r \, .\] \qed

\noindent We will say that the $\gsp (1)$-part of $\n \KK$ is {\bf small}
(at a point where $\KK=0$) if $\ |\nu f| < |\l_i|\ $ for all $i$. We will
say that the $\gsp (r)$-part is {\bf regular} if $\bar{L}_{\KK}$ is
invertible. This is the case if and only if the $\l_i \neq 0$.

\bc \label{Cor} Let $p\in M$ be a point where $\KK = 0$ and let the
$\gsp(1)$-part of $\n \KK$ be small (and therefore the $\gsp (r)$-part is
regular). Then the Hessian ${\rm Hess}_f$ has $r$ positive and $r$
negative eigenvalues, each of double multiplicity. \ec

We recall that all the {\bf known non-flat homogeneous \qk s} fall into
two classes: the Wolf spaces~\cite{Wolf1965} and the Alekseevsky
spaces~\cite{Alekseevsky1975,deWit:1992nm,Cortes1996}. The Wolf spaces are
symmetric spaces of positive scalar curvature. Their isometry group is
compact. The Alekseevsky spaces are precisely the homogeneous \qk s of
negative scalar curvature which admit an $\bR$-splittable simply
transitive solvable group of isometries. This class contains the
non-compact duals of the Wolf spaces, which are symmetric spaces of
negative scalar curvature, together with 3 series of non-symmetric \qk s.
The following result characterizes the symmetric \qk s.

\bt \cite{AlekseevskyC1997} A homogeneous \qk\,    is symmetric if and
only if it admits a smooth compact quotient by a discrete group of
isometries. \et

\bt Let $(M = G/H,g,Q)$ be one of the  known non-flat homogeneous
quaternionic-K\"{a}hler manifolds. Then there exists a Killing vector field
$\KK$ vanishing at a point $p\in M$ such that the Hessian ${\rm Hess}_f$
has split signature at $p$. \et

\pf By Corollary \ref{Cor} it is sufficient to prove that the isotropy Lie
algebra $\gh$ contains a vector with small $\gsp (1)$-part and regular
$\gsp (r)$-part. For the symmetric \qk s the isotropy Lie algebra splits
as a direct sum of ideals:
 \be \label{splitEqu}
  \gh = \gsp (1) \oplus
\gh'\, ,\quad \gh' \subset \gsp (r)\, .
 \ee
(For most of the symmetric \qk s $\gh' = [\gh' ,\gh' ]$. The only
exception is when $G$ is of type $A_{n+1}$, with $\gh' = \gu(n)$ and
$[\gu(n),\gu(n)] = \gsu(n)$). Let $\gt \subset \gh'$ be a Cartan
subalgebra and $T\in \gt$ a regular element. Then $T$ has regular $\gsp
(r)$-part (has only non-zero eigenvalues under the isotropy
representation) since the isotropy representation of $\gh'$ has no
trivial submodule. By adding a small vector from $\gsp (1)$ we obtain a
Killing vector with the desired properties.

The non-symmetric case is more involved since the isotropy group $\gh
\subset \gsp (1) \oplus \gsp (r)$ does not admit a splitting of the type
(\ref{splitEqu}). The isometry and isotropy groups of these spaces were
found in~\cite{deWit:1993wf} (see also summary in~\cite{deWit:1995tf}). We
use the description of the Alekseevsky spaces given in~\cite{Cortes2000}.
This does not use the \textit{full} isometry group. The metric-preserving
group in the centralizer of the Clifford algebra, which consists of the
antisymmetric matrices commuting with the gamma matrices, is not included.
This group is part of the full isometry and the full isotropy group.

Let $M = G/H$ be an Alekseevsky space of dimension $4r$. The Lie group
$G= G(\Pi )$ is defined by a $\gspin (V)$-equivariant map $\Pi : \wedge^2W
\ra V$, where $V$ is a pseudo-Euclidean vector space and $W$ is a module
for the even Clifford algebra $C\! \ell^0(V)$. The isotropy Lie algebra
has the form\footnote{The $q$ here is in agreement with the notation
in~\cite{Alekseevsky1975,Cortes1996,deWit:1992nm,deWit:1993wf,deWit:1995tf}.
The $P$ or $\dot P$ values in these papers determine the choice of the
module $W$. In~\cite{Cortes2000} a generalization is made to homogeneous
spaces of non-positive signature. This is reflected in the extra
parameter $p$, which is 0 in the positive-signature case.}
\begin{equation}
  \gh = \gso (3) \oplus \gso (p,q+3)\,,
 \label{HinAl}
\end{equation}
where $(p+3,q+3)$ is the signature of $V$. Note that the $\gso (3)$ here
is not the one that acts as the quaternionic structure, which we
consistently denote as $\gsp(1)$. We denote by $\pi : \gh \ra \gsp (r)$
the $\gsp (r)$-projection of $\gh$. It is faithful and has parts from
both subalgebras of (\ref{HinAl}). The $\gsp (1)$-projection of $\gh$ has
kernel $\gso (p,q+3)$ and defines an isomorphism $\gso (3) \ra \gsp (1) =
Q$. The isotropy module splits under $\pi (\gh) \cong \gso (3) \oplus
\gso (p,q+3)$ as follows
 \be \label{pidecomp}
  \bC^2 \oplus \bC^2 \otimes
\bR^{p,q+3} \oplus W\, .
 \ee
Here $\gso (3)$ acts on $\bC^2 = \bH$ by the standard representation of
$\gsu (2) \cong \gso (3)$ commuting with the quaternionic structure and
$\gso (p,q+3)$ acts trivially on $\bC^2$ and in standard way on
$\bR^{p,q+3}$. The action on the $C\! \ell^0(V)$-module $W$ is induced by
the inclusion
\[ \gso (3) \oplus \gso (p,q+3) \subset
\gso (V) = \gso (p+3,q+3) \cong \gspin (V) \subset C\! \ell^0(V)\, .\]
{}From this description we see that the isotropy module of $\pi (\gh)$
has no trivial submodules. This proves that $\gh$ contains elements with
regular $\gsp (r)$-part. Also it is easy to see that the $\gsp (1)$-part
of such an element has to be non-zero (due to the submodule $\bC^2$) and
can be chosen to be arbitrarily small. \qed

For the symmetric \qk s we can prove the existence of Killing vector
fields $\KK$ vanishing at a point $p\in M$ such that ${\rm Hess}_f$ is
definite at $p$ (non-degenerate local extremum of $f$).

\bt Let $(M = G/H,g,Q)$ be a symmetric \qk\, of reduced scalar curvature
$\nu$. If $\nu > 0$ (respectively, $\nu < 0$), there exists a Killing
vector field $\KK$ vanishing at a point $p\in M$ such that the Hessian
${\rm Hess}_f$ is negative (respectively, positive) definite, i.e.\ the
\namef\  $f$ has a non-degenerate local maximum (respectively, minimum) at
$p$. \et

\pf It follows from the decomposition \ref{splitEqu} that there exist
Killing vector fields $\KK$ vanishing at a point $p\in M$ with zero $\gsp
(r)$-part at $p$. For such a field $\KK$ the Hesse operator of $f$ at $p$
is given by ${\rm H}_f = - \nu f\, {\rm id}$. So ${\rm H}_f < 0$ if $\nu
>0$ and ${\rm H}_f > 0$ if $\nu < 0$. (The same is true for any Killing
vector with sufficiently small $\gsp (r)$-part.) \qed

\noindent
Note that for the non-symmetric Alekseevsky spaces there are no
non-zero Killing vector fields in the isotropy Lie algebra with zero
$\gsp (r)$-part. Nevertheless we can prove the following result.

\bt Let $(M = G/H,g,Q)$ be an Alekseevsky space. It is defined by a
$\gspin (V)$-equivariant map $\Pi : \wedge^2 W \ra V$, where $V =
\bR^{p+3,q+3}$ and $W$ is a $C\! \ell^0(V)$-module. Then there exists a
Killing vector field $\KK$ vanishing at a point $p\in M$ such that the
Hessian ${\rm Hess}_f$ is positive semi-definite at $p$. More precisely,
the spectrum of ${\rm Hess}_f$ consists of three eigenvalues: $\l := -
\nu f>0$, of multiplicity $4(r-p-q+2)$, the eigenvalue $2\l$, of
multiplicity $2(p+q+4)$, and the eigenvalue $0$, of multiplicity
$2(p+q+4)$. \et

\pf It is sufficient to choose $\KK \in \gso (3) \subset \gh = \gso (3)
\oplus \gso (p,q+3)$. At the canonical base point we can assume without
loss of generality that $P = \e J_1$ (and hence $f = \e^2$). Then the
Hesse operator at the canonical base point is ${\rm H}_f = -\nu f\, {\rm
id} + S = -\nu \e^2 \, {\rm id}+ S = - \e\, {\rm id}  + S$, where $S = \e
J_1\bar{L}_{\KK}$. Recall that the decomposition (\ref{pidecomp})  of the
isotropy module splits under the $\gsp (r)$-projection $\pi (\gh )$. With
respect to that decomposition $S$ acts trivially on $W$ and acts only on
the first factor $\bC^2$ of $\,\bC^2 \oplus \bC^2 \otimes \bR^{p,q+3} =
\bC^2 \otimes \bR^{p,q+4}\,$ with eigenvalues $\pm \e$ of double
multiplicity. This shows that ${\rm H}_f$ has eigenvalues $-\e$ (of
multiplicity $\dim W$), $-2\e$ (of multiplicity $2r -\dim W/2$) and $0$
(of multiplicity $2r -\dim W/2$). \qed

In the next sections we want to extend our discussion to the manifolds
which are allowed targets for the scalars of $5$-dimensional supergravity
theories. The most general such manifold for a theory with $r$
hypermultiplets and $n$ vector multiplets is a product $M \times N$ of a
\qk\, of dimension $4r$ and a {\it very special manifold} $N$ of
dimension~$n$.
\section{Very special real manifolds} \label{ss:vspR}
The geometry connected to vector multiplets in 5 dimensions was uncovered
in an old beautiful paper~\cite{Gunaydin:1984bi}. The manifolds were
placed in the context of the family of special geometries
in~\cite{brokensi}.

A {\bf very special manifold} is a connected immersed hypersurface $N
\hookrightarrow \{ C = 1\} \subset \bR^{n+1}$ defined by a homogeneous
cubic polynomial $C$ which is non-singular on a neighborhood of the image
of the immersion. For simplicity of our exposition, we assume, without
loss of generality, that $N \subset \bR^{n+1}$ is embedded. Then we do
not need to distinguish between points of $N$ and their images in
$\bR^{n+1}$. The radial vector field $\xi$ defined by $\xi (p) = p$ is
always transversal to the hypersurface $N$. It gives rise to a
pseudo-Riemannian metric $g = g_N$ and to a torsionfree connection $D$ on
$N$. They are defined by the formula\footnote{The factor $2/3$ is
introduced for consistency of the notation with the supergravity action.
It guarantees that the corresponding scalars have the same normalization
in the kinetic energy as the scalars of the quaternionic manifold. The
translation of these formulae to the familiar supergravity language will
be given at the end of this section.}
 \be \label{GWEqu}
 \partial_{\XX}\YY = D_{\XX}\YY + \ft23 g(\XX,\YY)\xi \, .
 \ee
Here $\XX$ and $\YY$ are tangent to $N$ and $\partial$ denotes the
canonical connection of $\bR^{n+1}$. Usually one assumes that the metric
is positive definite.

Let us denote by $C(\cdot , \cdot ,\cdot )$ the completely symmetric
trilinear form whose associated cubic form is $C = C(\cdot )$. They are
related by polarization: $C(\XX,\XX,\XX) = C(\XX)$.
 \bp
The metric $g$ is related to the Hessian of $C$ by
 \be
  g =  -\ft{1}{2}{\rm Hess}_C|N \,.\label{gHessEqu}
 \ee
More explicitly, for all $\XX,\YY\in T_pN$:
 \be \label{gEqu}
  g(\XX,\YY) = - 3C(p,\XX,\YY) \, .
 \ee \ep

\pf Equation (\ref{gHessEqu}) easily implies the explicit formula
(\ref{gEqu}) since $C$ is a homogeneous function of degree $3$. Let $\XX$
and $\YY$ be vector fields tangent to $N$. We can extend them (locally) to
vector fields in the ambient space $\bR^{n+1}$ satisfying $\XX C = \YY C =
0$. Hence, using the homogeneity of $C$ we obtain at a point $p \in \{ C
= 1\}$:
 \be \label{HessgEqu}
  {\rm Hess}_C(\XX,\YY) = \XX\YY C - \partial_{\XX}\YY C = 0 -
3C(p,p,\partial_{\XX}\YY) = - 3C(p,p,\partial_{\XX}\YY)  \, .
   \ee
The right hand side of (\ref{HessgEqu}) equals $-2 g(\XX,\YY)$ because the
linear form $C(p,p, \cdot )$ vanishes precisely on the tangent space of
$N$ at $p$ and equals $1$ on $p = \xi (p)$. \qed

\noindent
We define a $(1,2)$-tensor $S^C$ on $N$ by
 \be \label{SCEqu}
  g(S^C_{\XX}\YY,\ZZ)
= \ft32 C(\XX,\YY,\ZZ)\quad \mbox{for all} \quad \XX,\YY,\ZZ \in T_pN\, .
 \ee

\bl \label{LCLemma} The Levi-Civita connection of $g$ is given by
 \be
 \nabla = D - S^C\, .
 \ee \el

\pf The connection  $\nabla = D - S^C$ is torsionfree because $D$ is
torsionfree and $S^C_{\XX}\YY = S^C_{\YY}\XX$. We check that it is a
metric connection. Let $\XX,\YY,\ZZ$ be tangent vectors to $N$ (locally)
extended to vector fields in $\bR^{n+1}$ satisfying $\XX C = \YY C = \ZZ
C = 0$. Then we compute
 \bean
 (\nabla_{\XX} g) (\YY,\ZZ) &=& \XX g(\YY,\ZZ) - g(\n_{\XX}\YY,\ZZ) - g(\YY,\n_{\XX}\ZZ)\\
&=& \XX g(\YY,\ZZ) - g(D_{\XX}\YY,\ZZ) - g(\YY,D_{\XX}\ZZ) + g(S^C_{\XX}\YY,\ZZ) + g(\YY,S^C_{\XX}\ZZ)
\\
&\stackrel{(\ref{GWEqu})}{=}& \XX g(\YY,\ZZ)
- g(\partial_{\XX}\YY,\ZZ) - g(\YY,\partial_{\XX}\ZZ)
 + g(S^C_{\XX}\YY,\ZZ) + g(\YY,S^C_{\XX}\ZZ) \\
&=& (\partial_{\XX} g)(\YY,\ZZ) + 2C(\XX,\YY,\ZZ)\\
&\stackrel{(\ref{gHessEqu})}{=}& -\ft{1}{3}(\partial_{\XX} {\rm
Hess}_C)(\YY,\ZZ)
+ 2C(\XX,\YY,\ZZ)\\
&=&   -\ft{1}{3}(\partial^3C)(\XX,\YY,\ZZ) + 3C(\XX,\YY,\ZZ)\\
&=& -3C(\XX,\YY,\ZZ)- 3C(\XX,\YY,\ZZ) = 0\,.
 \eean
\qed

It is customary to denote the standard coordinates of $\bR^{n+1}$ by
$h^I$, with $I = 0,1,\ldots, n$, and the cubic polynomial is then
$C=C_{IJK}h^I h^J h^K$. The conjugate coordinates are
$h_I:=C_{IJK}h^Jh^K$. One may choose local coordinates $\phi ^x$ with
$x=1,\ldots ,n$ on the hypersurface $N$. Vector fields tangent to $N$ are
those $\YY$'s for which
\begin{equation}
 \YY=\YY^I\partial_I=\YY^x\partial_x\,,\qquad \rightarrow \qquad
  \YY^I=\YY^x(\partial_x h^I)
 \,.
 \label{YtangentN}
\end{equation}
The equation (\ref{GWEqu}) is then the decomposition of the derivative
\begin{equation}
  \partial_x\left[\YY^y(\partial_y h^I)\right]
   = (D_x \YY^y) (\partial_y h^I) + h^I g_{xy} \YY^y\,,
 \label{youreqn}
\end{equation}
and the orthogonality of $h^I$ and $\partial_y h^I$ implies that this
defines the metric $g_{xy}$. The lemma~\ref{LCLemma} corresponds to the
equation (where the semi-colon indicates covariant differentiation w.r.t.
$\phi ^x$ using Christoffel connection calculated from the metric
$g_{xy}$)~\cite{Gunaydin:1984bi}
\begin{equation}
   (\partial_y h^I)_{;x}
   =-\sqrt{\ft23}T_{xy}^z (\partial_z h^I) + h^I g_{xy}\,,\qquad
    T_{xyz} := \left( \ft32\right) ^{3/2} C_{IJK}
             (\partial_xh^I)(\partial_yh^J)(\partial_z h^K)\,,
 \label{eqnGST}
\end{equation}
such that the derivative $D$ on a vector tangent to the hypersurface
corresponds to
\begin{equation}
   D_x\YY^y=\YY^y{}_{;x}-\sqrt{\ft23}T_{xz}^y\YY^z\,.
 \label{Dincoord}
\end{equation}

\section{The dressed moment map}\label{ss:dressedP}
Let $(M,g_M,Q)$ be a \qk\, of dimension $4r$ and $(N,g_N)$ a very special
manifold of dimension $n$, $N\subset \bR^{n+1}$. We denote by $\pi_M$ and
$\pi_N$ the projections of the product $M\times N$ and by  $g =
\pi^*_Mg_M + \pi^*_Ng_N$ the product metric. We assume that we are given
a Lie algebra $\g$ spanned by $n+1$ Killing vector fields $\KK_I$, $I =
0,1,\ldots, n$, on $M$, with the corresponding moment maps being $P_I :
M\ra Q$. We define the {\bf dressed moment map} $P: M\times N\ra Q\,$ by
 \be
  P := P^\a J_\a :=  h^IP_I\qquad \mbox{or}\qquad P_{XY}:= P^\alpha J_{\alpha\,XY}\, .
 \ee
 It is a section of the bundle $\pi_M^*Q$ over $M\times N$, where
$\pi_M : M\times N \ra M$ is the projection. Let us also
define\footnote{In sections~\ref{ss:qKmap} and~\ref{ss:Hessianq} we
considered one isometry whose Killing vector we denoted by $\KK$ and
whose moment map we denoted by $P$. This can be considered in the context
of this section as the case of a trivial very special real manifold, i.e.
$n=0$, with $h^I$ having only the component $h^0=1$.}
 \be
 \KK := h^I\KK_I\, .
\ee It is an $N$-dependent vector field on $M$, a section of $\pi_M^*
TM$. We want to study the gradient flow generated by the {\bf  \namef}
function $f := P^2$.

\bp The covariant derivative of the dressed moment map $P$ is given by
\be \label{nablaPgeneral}
 \n P = \rmd h^I\otimes P_I +  h^I \n P_I = \rmd
h^I\otimes P_I + \ft{1}{2} \rho_\a (\cdot , \KK) \otimes J_\a \, .
 \ee
 The differential of the  \namef\  is
\be \rmd f = 2\langle \n P, P\rangle = 2 \langle P_I,P\rangle \rmd h^I +
P^\a \rho_\a (\cdot , \KK) \, .
 \ee
  The gradient of the \namef\  is
 \be {\rm
grad}\, f = 2 \langle P_I,P\rangle {\rm grad}\, h^I + P\KK\, .
 \ee \ep

\pf This follows easily from Proposition~\ref{firstProp} applied to the
$P_I$. \qed

\bc \label{CritCor} The set of critical points of $P$
is~\cite{Ceresole:2001wi}
 \be
  {\rm Crit} (P) = \{  \rmd h^I\otimes P_I =
0\} \cap \{ \KK = 0\} = \{ P_I = h_IP\} \cap \{ \KK = 0\} \, .
 \ee
 The set of critical points of the \namef\  $f$ is
\bea
 {\rm Crit}(f) &=& \Bigl\{  \langle P,P_I \rangle \rmd h^I = 0\Bigr\} \cap \Bigl(\{ \KK
= 0\} \cup \{ f = 0\}\Bigr)
\nonumber\\
\\ \nonumber
&=& \Bigl\{ \langle P,P_I\rangle = fh_I\Bigr\} \cap \Bigl(\{ \KK = 0\}
\cup \{ f = 0\}\Bigr) \, .
 \eea
 \ec
In the supergravity context, critical points are points with constant
scalars for solutions that preserve supersymmetry. The preservation of
supersymmetry imposes for the $N$-sector the condition $P_I=h_I P$, i.e.
it are the critical points of $P$.

\section{The Hessian of the \namef\ }\label{ss:Hessfull}
In this section we carry over the calculations from section
\ref{ss:Hessianq} to case of the dressed moment map.

\bl \label{2ndderLemma} The second covariant derivative of the dressed
moment map is given by:
 \bea
  \n^2_{\XX,\YY}P &=&  {\rm Hess}_{h^I}(\XX,\YY)P_I +\ft12
 \rho_\a (\XX ,\KK_I)\YY(h^I) J_\a + \ft12 \XX(h^I)\rho_\a (\YY ,\KK_I)J_\a \nonumber\\
& & +\ \ft1{2} h^I\rho_\a (\YY , \n_{\XX} \KK_I)J_\a\, .
 \eea
The Hessian of the \namef\  is:
 \bea
   {\rm Hess}_f (\XX,\YY) &=& 2 \langle P,P_I\rangle {\rm
Hess}_{h^I}(\XX,\YY) +
 P^\a \rho_\a (\XX ,\KK_I)\YY(h^I)\nonumber \\
& & + P^\a \XX(h^I) \rho_\a (\YY ,\KK_I)
+  P^\a h^I\rho_\a (\YY, \n_{\XX} \KK_I)\nonumber \\
& & +\ 2 \langle \n_{\XX} P ,\n_{\YY} P\rangle.
 \eea
\el

\pf Let us compute from (\ref{nablaPgeneral}): \bea \n^2 P =  (\n \rmd
h^I)\ot P_I +  \n P_I \ot \rmd h^I + \rmd h^I \ot \n P_I + h^I\n^2P_I\, .
\eea This finishes the proof in view of Lemma \ref{nabla2PLemma}. \qed

\noindent We put $\bar{L}_{\KK} :=  h^I\bar{L}_{\KK_I}$.
 \bt
The Hessian of the \namef\  is given by
 \bean
  {\rm Hess}_f &=& \ft43f \pi^*_Ng_N +2 \langle P, P_I\rangle \rmd h^IS^C
  \\ &&
  +  \rmd h^I \ot gP\KK_I + gP\KK_I \ot \rmd h^I
  \\ &&
   +\pi^*_Mg_M (-\nu f {\rm id} + S) + \n P\ot \n P \, ,
 \eean
where $S^C$ is the $(1,2)$-tensor defined by (\ref{SCEqu}) and $S$ is the
symmetric operator $S:= P\bar{L}_{\KK}$, $gA = g(A \cdot , \cdot )$ for an
endomorphism $A$ and $g\XX = g(\XX,\cdot )$ for a vector $\XX$.
 \et

\pf Using  Lemma \ref{2ndderLemma} we obtain
 \bean
{\rm Hess}_f &=& 2 \langle P,P_I\rangle {\rm Hess}_{h^I}
+ \rmd h^I \ot gP\KK_I + gP \KK_I\ot \rmd h^I+ h^I gP\n \KK_I\\
& & +\  2 \langle \n P \ot \n P\rangle \, .
 \eean
The decomposition (see (\ref{decompNK}))
 \be \n \KK_I = \nu P_I + \bar{L}_{\KK_I}
 \ee shows that
  \be
h^I gP\n \KK_I = - \nu f \pi^*_Mg_M + gS\, .
 \ee

To simplify the first term we need to calculate the Hessian ${\rm
Hess}_{h^I}$ of the function $h^I|N$ with respect to the Levi-Civita
connection $\n$ of $N$. Of course, the Hessian $\partial^2h^I$ of the
linear function $h^I$ with respect to the standard connection of
$\bR^{n+1}$ is zero. This means that $\XX\YY h^I = (\partial_{\XX}\YY)h^I$
for all vector fields $\XX$ and $\YY$. So, using (\ref{GWEqu}) and Lemma
\ref{LCLemma}, we get for all vector fields $\XX$ and $\YY$ tangent to
$N$:
 \bean
 {\rm Hess}_{h^I}(\XX,\YY) &=&  \XX \YY h^I - (\n_{\XX}\YY)h^I =
(\partial_{\XX}\YY - \n_{\XX}\YY)h^I \\
&=&  (D_{\XX}\YY + \ft23g_N(\XX,\YY)\xi - \n_{\XX}\YY)h^I =
\ft23g_N(\XX,\YY) h^I + (S^C_{\XX}\YY)h^I\,.
 \eean
  \qed

\noindent
In components the Hessian of $h^I$ follows from (\ref{eqnGST}).
The expression $gP\KK_I$ is the one-form with components
\begin{equation}
  (gP\KK_I)_X=-P_{XY}\KK^Y_I\,.
 \label{defcalJ}
\end{equation}
On the other hand, the components of $S$ (symmetric and traceless) are
\begin{equation}
  S_{XY}= P^\alpha J_{\alpha X}{}^Z\left(\bar L_{\KK} \right)_{ZY}\,.
 \label{compS}
\end{equation}
The result of the theorem can therefore be written as
\begin{eqnarray}
 \partial_x\partial_y f& = & \ft43\left( fg_{xy}-
 \sqrt{\ft23}P^\alpha P^\alpha _I T_{xy}^z\partial_zh^I\right),
 \nonumber\\
 \partial_x\partial_X f & = & -P_{XY}\KK^Y_I\partial_xh^I\,,
 \nonumber\\
 \partial_X\partial_Y f&=& -\nu g_{XY} f+S_{XY}\,.
 \label{ddfcomp}
\end{eqnarray}

\bc At a critical point of $P$ the Hesse operator of the \namef, defined
as
\begin{equation}
  {\rm H}_f(\cdot ,g\cdot )={\rm Hess}_f(\cdot ,\cdot )\,,
 \label{defHesse}
\end{equation}
is given by
 \be {\rm H}_f  = \ft43f \pi_{N*} + \rmd h^I \otimes P\KK_I +
gP\KK_I\otimes {\rm grad}\, h^I + (- \nu f {\rm id} +  S)\circ \pi_{M*}\,.
 \ee
Here $\pi_{M*}$ and $\pi_{N*}$ are the differentials of the canonical
projections.
 \ec

\pf It suffices to remark that at a critical point $\n P{=} 0$ and
$P_I\rmd h^I {=} 0$, see Corollary~\ref{CritCor}. \qed

\section{Conclusions}\label{ss:conclusions}

We have considered the properties of flows governed by (\ref{dxQ0}) and
(\ref{floweqns}). In particular, we have obtained a formula for the
Hessian matrix ${\cal U}$ defined in (\ref{defcalU}):
\begin{equation}
  {\cal U}:=\left.\frac{3}{2f}{\rm H}_f\right|_{\partial f=0}
  =\pmatrix{2\delta _x{}^y& +\frac1{W^2} (\partial_x\KK^Z)P_Z{}^Y\cr
  -\frac1{W^2} P_{XZ}\partial ^y\KK^Z
  &-\nu \frac32\delta _X{}^Y+ \frac{1}{W^2}P_X{}^Z
  \left(\bar L_{\KK}\right)_{Z}{}^Y}\,,
 \label{calUres}
\end{equation}
where the first entries are for the vector multiplets and the second for
the hypermultiplets. $\nu $ is the reduced scalar curvature, see
theorem~\ref{structureThm}, which is $-1$ in supergravity. $P$ and $\bar
L_{\KK}$ select respectively the $\gsp (1)$ and $\gsp(r)$ parts of the
dressed gauged isometry, defined by (\ref{defcalJ}) and
\begin{equation}
 {\cal D}_X \KK_Y=\nu P_{XY}+ \left(\bar  L_{\KK} \right)_{XY}\,.
 \label{vpro-defcalJL}
\end{equation}
Thus, in comparison with~\cite{Ceresole:2001wi,VanProeyen:2001ng}, $P$ is
$-{\cal J}$ and $\bar L_{\KK}$ is ${\cal L}$.

Note that if only vector multiplets are present, we have only the
upper-left entry of (\ref{calUres}), and thus only positive eigenvalues
(UV attractors)~\cite{KL,BC}. However, including the quaternionic-K\"{a}hler
manifold (hypermultiplets) opens the possibility of having negative
eigenvalues as well~\cite{DW,Ceresole:2001wi}. The formula
(\ref{calUres}) implies that such eigenvalues are only possible if either
the gauged isometries are mainly in the $\gsp(r)$ direction (${\bar
L}_{\KK}$ big enough) or one gauges generators that are not in the
isotropy group of the fixed point (such that $\KK_I^X\neq 0$ and the
off-diagonal elements are non-zero).

The lower-right $4r\times 4r$ part of (\ref{calUres}) has eigenvalues
\begin{equation}
-\nu \left(\ft32 +\lambda_1 \,,\, \ft32 +\lambda_1\,,\, \ft32-\lambda_1
\,,\,
  \ft32-\lambda_1\,,\, \ft32+\lambda_2\,,\,\ft32 +\lambda_2\,,\,
  \ft32-\lambda_2\,,\,\ft32-\lambda_2\,, \ldots,\,\ft32-\lambda_r\right) .
 \label{eigenlr}
\end{equation}

In various situations with trivial very special real manifold, we have
found more detailed results on the structure of the eigenvalues, and the
number of possible critical points. In particular, for complete
quaternionic-K\"{a}hler manifolds of non-positive sectional curvature (which
include the  locally symmetric quaternionic-K\"{a}hler manifolds of negative
scalar curvature) we find that the fixed point set, if non-empty, is
either a point or a connected totally geodesic submanifold of non-zero
dimension. Note that here, the fixed point set is non-empty if and only
if $\KK$ is a compact generator, i.e.\ if the closure of the group
generated by $\KK$ is compact. For homogeneous quaternionic-K\"{a}hler
manifolds we prove the existence of Killing fields $\KK$ such that the
spectrum of $\,\cal U$ has some specific UV/IR-properties, for example:
\begin{description}
\item{(i)} For all known  quaternionic-K\"{a}hler manifolds
we exhibit a Killing vector field $\KK$ vanishing at a point $p\in M$
such that  $\,\cal U$ has split signature at $p$.
\item{(ii)} For the symmetric quaternionic-K\"{a}hler manifolds
of positive scalar curvature we find a Killing vector field $\KK$
vanishing at a point $p\in M$ such that $\,\cal U$ is negative definite,
i.e.\ the \namef\  $f$ has a non-degenerate local maximum at $p$.
\item{(iii)} For the symmetric quaternionic-K\"{a}hler manifolds
of negative scalar curvature we find a Killing vector field $\KK$
vanishing at a point $p\in M$ such that $\,\cal U$ is positive definite,
i.e.\ the \namef\  $f$ has a non-degenerate local minimum at $p$.
\item{(iv)} For the known non-symmetric homogeneous
quaternionic-K\"{a}hler manifolds (Alekseevsky spaces), which have negative
scalar curvature,   we find a Killing vector field $\KK$ vanishing at a
point $p\in M$ such that $\cal U$ is positive semi-definite, i.e.\ the
\namef\  $f$ has a degenerate local minimum at $p$. Moreover we calculate
the eigenvalues of $\,\cal U$.
\end{description}
We remark that supergravity selects negative scalar curvature ($\nu =-1$),
so situation (ii) never occurs. We thus have either complete UV fixed
points as in (iii), or zero directions as in (iv), or split signatures as
in (i).

These results will be useful for the investigation of possibilities for
flow lines from IR to UV critical points, suitable for AdS/CFT dual
pairs, or from IR to IR critical points, relevant for the investigation
of possible smooth supersymmetric domain-wall solutions. We also expect
that the main results of this paper can also be applied to other
dimensions where quaternionic-K\"{a}hler manifolds occur, e.g. 4-dimensional
$N=2$ supergravity theories.

\newpage
\appendix
\section{Remarks on the notation}
Throughout this paper $Sp(n)$ denotes the compact real form of the
symplectic group in $2n$ variables, which is sometimes denoted as
$USp(2n)$. In particular $Sp(1) = USp(2) = SU(2)$.

Generically, ${\cal D}$ is a derivative that is covariant under all
existing local symmetries. In other words, it is a connection in all
bundles that are active on the field on which ${\cal D}$ acts. This means
that ${\cal D}$ is an extension of the Levi-Civita connection (when
acting on tangent vectors) adding a term $-$ (gauge field one-form)
$\times $ gauge transformation, for any gauge transformation under which
the object transforms. E.g. the latter gives rise to the last term
in~(\ref{calD}), containing the $\gsp(1)$ gauge field $\omega _\alpha $
as it acts there on an object that transforms under $\gsp(1)$. The
equation~(\ref{calD}) determines $\omega _\alpha$ and gives our convention
for the $\gsp(1)$ transformation on triplets.

The curvature tensor of the Levi-Civita connection $\n$ is defined by
(denoting vector fields $\XX$ in a coordinate basis as $\XX^\Lambda
\partial_\Lambda $)
\begin{eqnarray}
 \nabla_{\XX}\nabla_{\YY} \ZZ-\nabla_{\YY}\nabla_{\XX} \ZZ& = & \nabla_{[\XX,\YY]}\ZZ +R(\XX,\YY)\ZZ\,,\nonumber\\
 \langle R(\XX,\YY)\ZZ,\WW\rangle  & = & R^\Upsilon{}_{\Xi\Lambda\Sigma}
 \XX^\Lambda \YY^\Sigma \ZZ^\Xi \WW_\Upsilon\,.
 \label{defnR}
\end{eqnarray}
$\nabla ^2$ is defined as
\begin{equation}
  \nabla ^2_{\XX,\YY}
  \equiv \nabla_{\XX}\nabla_{\YY}- \nabla_{\nabla_{\XX}\YY}\,.
 \label{nabla2}
\end{equation}

The curvature of a connection ${\cal D}$ acting on tangent scalars in a
space whose components are denoted by indices $\Lambda , \Xi ,\ldots $ is
given by
\begin{equation}
  \left[ {\cal D}_\Lambda ,{\cal D}_\Xi \right] = -{\cal R}^a_{\Lambda \Xi
  }T_a
 \label{defCurvatures}
\end{equation}
for any gauge symmetry denoted by indices $a$, and whose action is
indicated here by $T_a$. See~(\ref{defcalR}) for the $\gsp(1)$ curvature.

Acting on a vector field $\XX$ with components $\XX^\Xi$ in a local
coordinate basis (neutral under other gauge transformations),
$\nabla_{\YY} \XX=\YY^\Lambda({\cal D}_\Lambda
\XX^\Sigma)\partial_\Sigma$, and the curvature components of the
Levi-Civita connection are given by
\begin{equation}
{\cal D}_\Lambda {\cal D}_\Sigma \XX^\Xi -{\cal D}_\Sigma {\cal
D}_\Lambda \XX^\Xi
= R^\Upsilon{}_{\Xi\Lambda\Sigma}\XX^\Xi\,.
 \label{nablanablac}
\end{equation}
The Ricci tensor and scalar curvature are
\begin{equation}
  Ric_{\Lambda \Sigma }=R^\Xi{}_{\Lambda\Xi\Sigma}\,, \qquad
  scal= g^{\Lambda \Sigma} Ric_{\Lambda \Sigma }\,.
 \label{Ricscal}
\end{equation}
\newpage

\providecommand{\href}[2]{#2}\begingroup\raggedright\endgroup

\end{document}